# Efficient Denial of Service Attack Detection in IoT using Kolmogorov-Arnold Networks

Oleksandr Kuznetsov [1,2*]

[1] Department of Theoretical and Applied Sciences, eCampus University, Via Isimbardi 10, Novedrate (CO), 22060, Italy

[2] Department of Intelligent Software Systems and Technologies, School of Computer Science and Artificial Intelligence, V.N. Karazin Kharkiv National University, 4 Svobody Sq., 61022 Kharkiv, Ukraine

*Corresponding author

Email: oleksandr.kuznetsov@uniecampus.it, kuznetsov@karazin.ua

https://orcid.org/0000-0003-2331-6326

**Abstract**: The proliferation of Internet of Things (IoT) devices has created a pressing need for efficient security solutions, particularly against Denial of Service (DoS) attacks. While existing detection approaches demonstrate high accuracy, they often require substantial computational resources, making them impractical for IoT deployment. This paper introduces a novel lightweight approach to DoS attack detection based on Kolmogorov-Arnold Networks (KANs). By leveraging spline-based transformations instead of traditional weight matrices, our solution achieves state-of-the-art detection performance while maintaining minimal resource requirements. Experimental evaluation on the CICIDS2017 dataset demonstrates 99.0% detection accuracy with only 0.19 MB memory footprint and 2.00 ms inference time per sample. Compared to existing solutions, KAN reduces memory requirements by up to 98% while maintaining competitive detection rates. The model's linear computational complexity ensures efficient scaling with input size, making it particularly suitable for large-scale IoT deployments. We provide comprehensive performance comparisons with recent approaches and demonstrate effectiveness across various DoS attack patterns. Our solution addresses the critical challenge of implementing sophisticated attack detection on resource-constrained devices, offering a practical approach to enhancing IoT security without compromising computational efficiency.

**Keywords**: Internet of Things security, denial of service attacks, Kolmogorov-Arnold Networks, intrusion detection systems, resource-efficient machine learning, edge computing security, spline-based neural networks, network traffic analysis

## I. INTRODUCTION

The rapid growth of Internet of Things (IoT) devices has created unprecedented security challenges [1,2]. Denial of Service (DoS) attacks pose a particular threat, targeting resource-constrained IoT devices that lack robust defense mechanisms [3,4]. While machine learning approaches have shown promise in detecting these attacks, they often require significant computational resources, making them impractical for IoT deployment [1,5].

This paper introduces a novel approach to DoS attack detection using Kolmogorov-Arnold Networks (KANs) [6]. KANs represent a fundamental advancement in neural network architecture, replacing traditional weight matrices with learnable spline functions. This architectural shift enables efficient pattern recognition with significantly fewer parameters than conventional neural networks.

Our approach addresses three critical challenges in IoT security:

- First, resource efficiency. IoT devices operate under strict computational and memory constraints. Our KAN-based solution requires only 0.19 MB of memory and achieves inference times of 2.00 ms per sample, making it suitable for deployment on resource-constrained devices;
- Second, detection accuracy. DoS attacks continue to evolve, requiring sophisticated detection mechanisms. Our approach achieves 99.0% accuracy while maintaining low false positive rates (1.6%), demonstrating robust detection capabilities across various attack patterns;
- Third, scalability. IoT networks can encompass thousands of devices, necessitating efficient detection solutions. The linear computational complexity of our approach ensures predictable scaling with network size.

The main contributions of this work include:

1. A lightweight DoS detection model based on KAN architecture, specifically optimized for IoT environments;
2. Comprehensive performance evaluation using the CICIDS2017 dataset, demonstrating competitive accuracy with minimal resource requirements;
3. Comparative analysis with state-of-the-art approaches, highlighting the efficiency-performance trade-offs;
4. Open-source implementation enabling reproducibility and further research.

Recent work in neural operators has demonstrated KANs' effectiveness across various domains, from physics simulations to time series analysis. Our work extends these advantages to IoT security, showing that sophisticated attack detection is possible without excessive computational overhead.

The rest of this paper is organized as follows. Section II reviews related work in IoT security and KAN applications. Section III provides background on KAN architecture and its theoretical foundations. Section IV details our methodology, followed by results and comparative analysis in Section V. Section VI discusses implications and limitations, with conclusions presented in Section VII.

**II. RELATED WORK**

Recent research in IoT intrusion detection has seen significant advances in both architectural approaches and detection performance. We organize this review around three key themes: deep learning architectures, feature optimization techniques, and resource-efficient solutions.

**Deep Learning Architectures**

Recent work has explored various deep learning architectures for intrusion detection. Awan et al. (2025) [7] proposed SecEdge, integrating transformer-based models with Graph Neural Networks (GNNs) for IoT security. While achieving 98.7% accuracy, their approach requires substantial computational resources. Similarly, Gamal et al. (2024) [8] developed an LSTM-RNN architecture for drone network protection, achieving 99.85% accuracy but requiring significant computational overhead.

Cherfi et al. (2024) [9] introduced an ALNS-CNN hybrid combining convolutional neural networks with adaptive large neighborhood search. Their approach achieved 99.85% accuracy on the CICIDS2017 dataset, though at the cost of high CPU utilization.

**Feature Optimization Techniques**

Several researchers have focused on optimizing feature selection and processing. Bikila and Čapek (2025) [5] proposed an Elastic Deep Autoencoder with Grey Wolf Optimizer (EDA-GWO) for feature extraction, achieving 99.87% accuracy. Their approach demonstrates excellent detection performance but requires complex optimization procedures.

Kumar et al. (2025) [10] developed NIDS-DA, focusing on non-functional feature separation for adversarial attack detection. Their approach achieved 99.97% accuracy with relatively low computational requirements (5,264 parameters), though specific hardware requirements weren't detailed.

**Resource-Efficient Solutions**

Recent work has increasingly emphasized resource efficiency for IoT deployment. Rajathi and Rukmani (2025) [11] introduced a Hybrid Learning Model (HLM) combining parametric and non-parametric classifiers. Their approach achieved 99.63% accuracy while attempting to minimize computational overhead through efficient feature selection.

Janati Idrissi et al. (2025) [12] investigated the impact of flow timeouts on model performance, highlighting the importance of efficient data processing in resource-constrained environments. Their work demonstrated that proper timeout configuration could significantly impact detection performance without increasing computational requirements.

**Research Gaps**

While existing approaches demonstrate impressive detection accuracy, several challenges remain unaddressed:

1. Most solutions require significant computational resources or complex preprocessing steps;
2. Few approaches provide comprehensive resource utilization metrics;
3. The trade-off between detection accuracy and resource efficiency is often not fully explored.

Our work addresses these gaps by introducing a KAN-based approach that achieves competitive detection performance while maintaining minimal resource requirements. Unlike previous solutions that focus primarily on detection accuracy, our approach explicitly considers computational efficiency and deployment practicality in IoT environments.

## III. BACKGROUND

KANs represent a fundamental shift in neural network architecture [6], inspired by Kolmogorov's universal approximation theorem [13,14]. The theorem states that any continuous multivariate function can be represented as:

$$f(x_1,...,x_d) = \sum_{q=1}^{2d+1} g_q\left(\sum_{p=1}^{d} \phi_{p,q}(x_p)\right),$$

where $\phi_{p,q}$ and $g_q$ are continuous univariate functions. KANs implement this theorem through a network architecture where, unlike traditional MLPs with fixed activation functions, the activation functions themselves are learnable through spline-based transformations.

In KANs, each edge transformation is represented by a cubic spline function [6]:

$$s(x) = \sum_{i=1}^{n} c_i B_i(x),$$

where $B_i(x)$ are basis spline functions and $c_i$ are learnable coefficients. This formulation enables KANs to approximate complex functions with fewer parameters than traditional neural networks.

The foundational work by Liu et al. (2024) [6] demonstrated that for a given accuracy target $\epsilon$, KANs require $O(\epsilon^{-d/r})$ parameters, compared to $O(\epsilon^{-2d/r})$ for MLPs, where $d$ is the input dimension and $r$ is the smoothness of the target function.

Recent applications have extended KANs to various domains. Abueidda et al. (2025) [15] developed DeepOKAN for solving complex engineering problems, while Wang et al. (2025) [16] introduced KINN for physics-informed solutions of partial differential equations. Danish and Grolinger (2025) [17] adapted KANs for time series analysis through KARN (Kolmogorov-Arnold Recurrent Network), defined as:

$$h_t = \sigma\left(\sum_{q=1}^{Q} g_q\left(\sum_{p=1}^{P} \phi_{p,q}(x_t^p) + \sum_{m=1}^{M} \psi_{m,q}(h_{t-1}^m)\right)\right),$$

where $h_t$ represents the hidden state at time $t$, and $\psi_{m,q}$ are additional spline functions for temporal dependencies.

The key advantages of KANs for IoT security applications include:

1. Parameter Efficiency: $O(n)$ scaling with input size;
2. Natural Regularization: Spline continuity constraints provide inherent regularization;
3. Interpretability: Edge transformations $s(x)$ can be directly visualized;
4. Computational Efficiency: Linear complexity in both training and inference.

These characteristics make KANs particularly suitable for IoT security applications, where efficient threat detection must be balanced against limited computational resources. Our work extends these advantages to DoS attack detection in IoT environments, demonstrating that KANs can achieve high detection accuracy while maintaining minimal resource requirements.

## IV. METHODOLOGY

The methodology section describes our lightweight approach to DoS attack detection using KANs. We detail the network architecture, data preprocessing steps, training procedure, and evaluation framework. Our method achieves high accuracy while maintaining minimal computational requirements suitable for IoT devices.

**A. Overview of the Proposed Approach**

Our approach combines KAN with carefully designed preprocessing steps to create a resource-efficient DoS detection system. KAN, based on Kolmogorov's universal approximation theorem [13,14], represents functions through nested compositions of simpler functions using spline operations. Unlike traditional neural networks that use weight matrices and nonlinear activations, KAN employs adaptive splines to model complex patterns while requiring significantly fewer parameters.

The proposed system consists of three main components:

1. A data preprocessing pipeline that handles missing values, removes outliers, and normalizes features;
2. A compact KAN architecture with three hidden layers optimized for IoT deployment;
3. A comprehensive evaluation framework that analyzes both detection accuracy and resource utilization.

Figure 1 presents the overall architecture of our system.

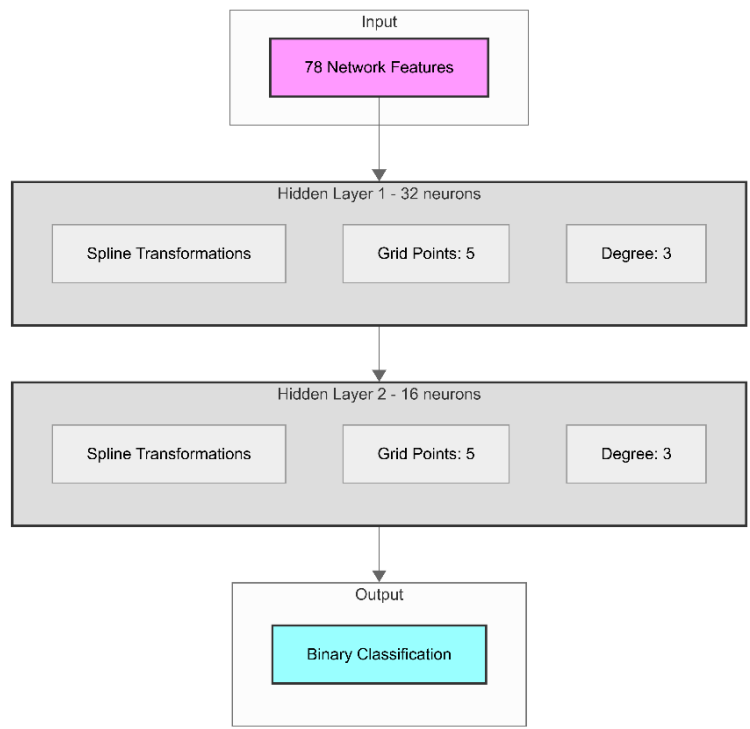

Figure 1: System Architecture Diagram

The key advantages of our approach include:

- Lightweight architecture with only 50K parameters (0.19 MB);
- Fast inference time of 2.00ms per sample;
- High detection accuracy (99.0%);
- Built-in visualization capabilities for network decision boundaries.

**B. Kolmogorov-Arnold Networks Architecture**

The core of our approach is a specialized KAN architecture optimized for IoT-based DoS detection. KAN implements Kolmogorov's superposition theorem through adaptive spline operations, offering several advantages over traditional neural networks for resource-constrained environments.

Our network consists of three main layers (Figure 1):

1. Input layer: Processes 78 network traffic features
2. Hidden layers: Two layers with 32 and 16 neurons respectively
3. Output layer: Single neuron for binary classification (attack/normal)

Each hidden layer employs:

- 5 grid points for spline discretization
- Cubic splines (degree 3) for function approximation
- Adaptive connection patterns between layers

The key innovation in our architecture lies in its compact design. By using spline-based transformations instead of traditional weight matrices, we achieve:

- 50,092 total parameters (42,336 trainable)
- 0.19 MB model size
- Efficient function approximation capability

Figure 2 shows the internal spline transformations for selected neurons, demonstrating how the network learns to separate normal and attack patterns through nonlinear feature transformations.

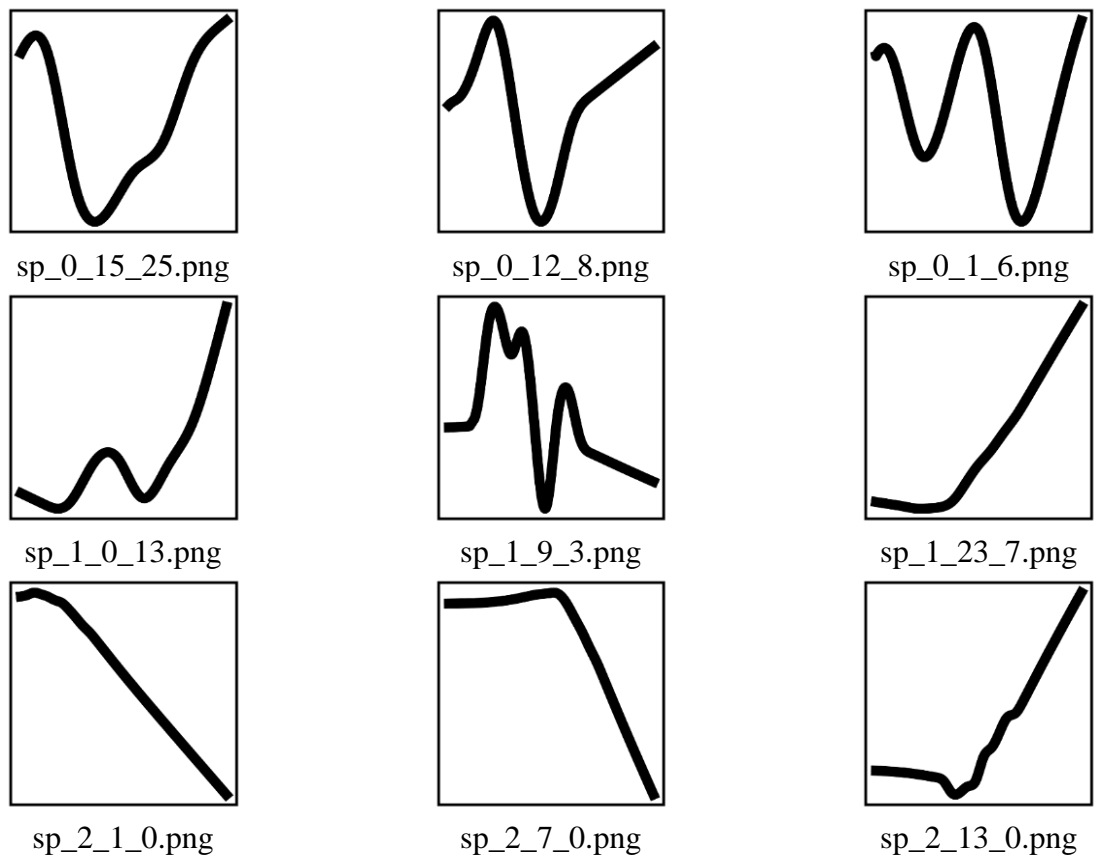

Figure 2: Selected Spline Visualizations

This architecture balances model capacity with computational efficiency, making it suitable for deployment on resource-constrained IoT devices while maintaining high detection accuracy.

**C. Dataset and Preprocessing**

Our approach uses the CICIDS2017 dataset, specifically the Wednesday traffic data containing various DoS attacks:

- DoS Hulk (231,073 samples);
- DoS GoldenEye (10,293 samples);
- DoS slowloris (5,796 samples);
- DoS Slowhttptest (5,499 samples);
- Heartbleed (11 samples).

The preprocessing pipeline implements these key steps:

1. Data Selection and Balancing:
    - Used maximum available balanced samples (231,073 per class);
    - Maintained attack/normal ratio 1:1;
    - Total dataset: 462,146 samples;
2. Feature Engineering:
    - Selected 78 network traffic features;
    - Applied outlier removal (3σ rule);
    - Used median imputation for missing values;
    - Implemented standardization (zero mean, unit variance);

The attack distribution and feature importance analysis are shown in Figures 3 and 4 respectively.

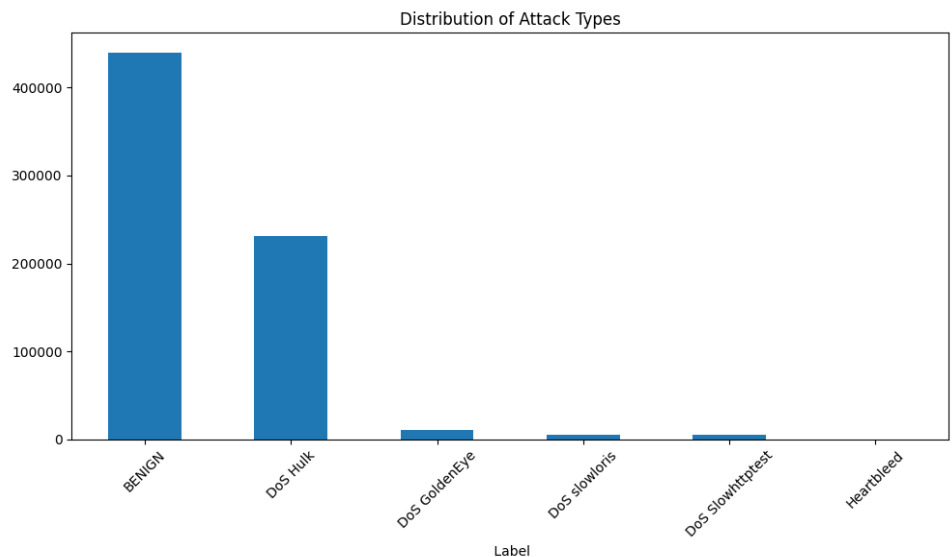

Figure 3: Attack Types Distribution

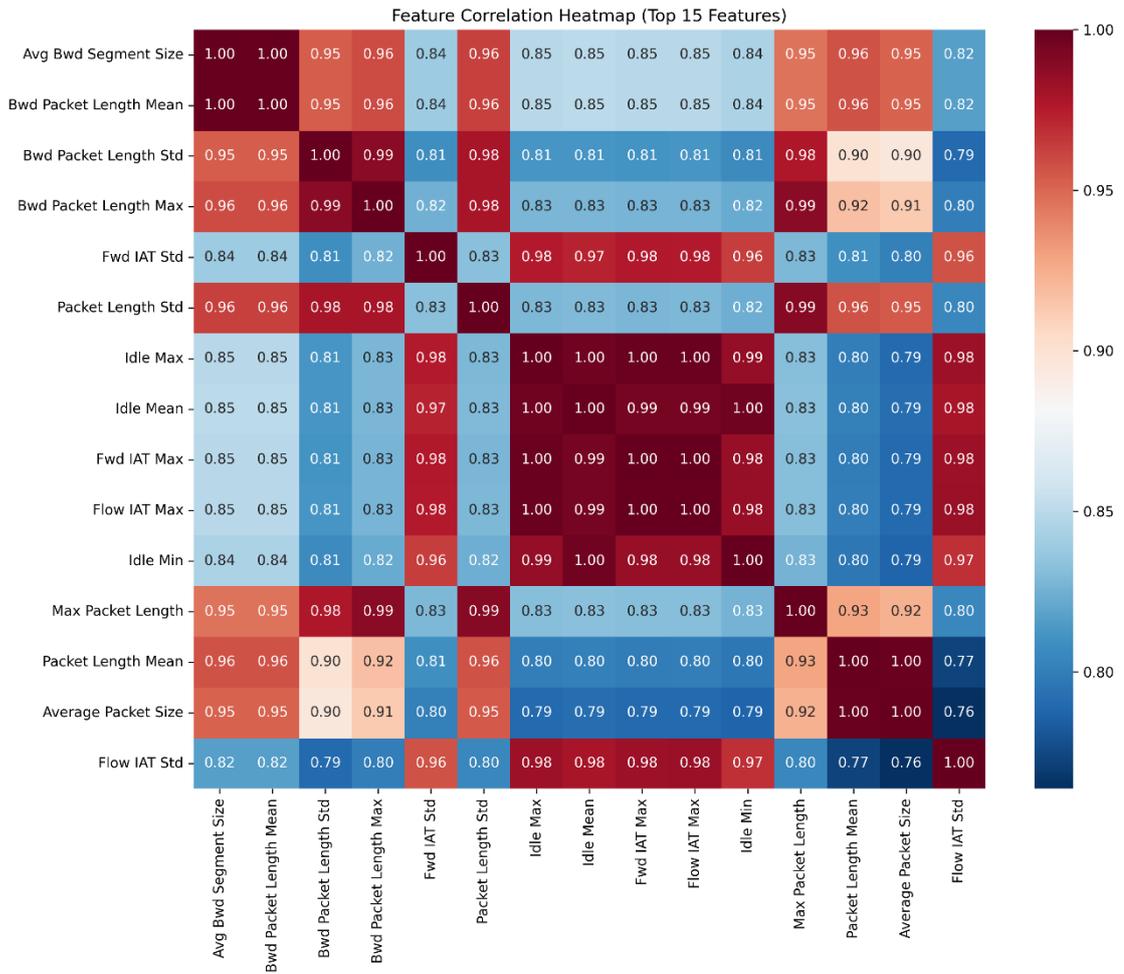

Fig. 4: Correlation Heatmap of Features (Top 15 features)

The dataset preparation code and detailed feature statistics are available in our GitHub repository under the data/ directory.

**D. Network Training Procedure**

We implemented an efficient training procedure optimized for IoT deployment constraints. The training process includes:

1. Optimization Settings:
    - Adam optimizer with learning rate 0.001;
    - Binary cross-entropy loss function;
    - Batch size: 100 samples;
    - Maximum epochs: 200;

2. Model Configuration:
    - Input layer: 78 features;
    - Hidden layers: [32, 16] neurons;
    - Output layer: 1 neuron (binary classification);
    - Spline parameters: grid points = 5, degree = 3;

3. Training Dynamics. Figure 5 demonstrates the training process performance:
    o Fig. 5a shows the loss convergence for both training and test sets;
    o Fig. 5b presents accuracy metrics over training epochs.

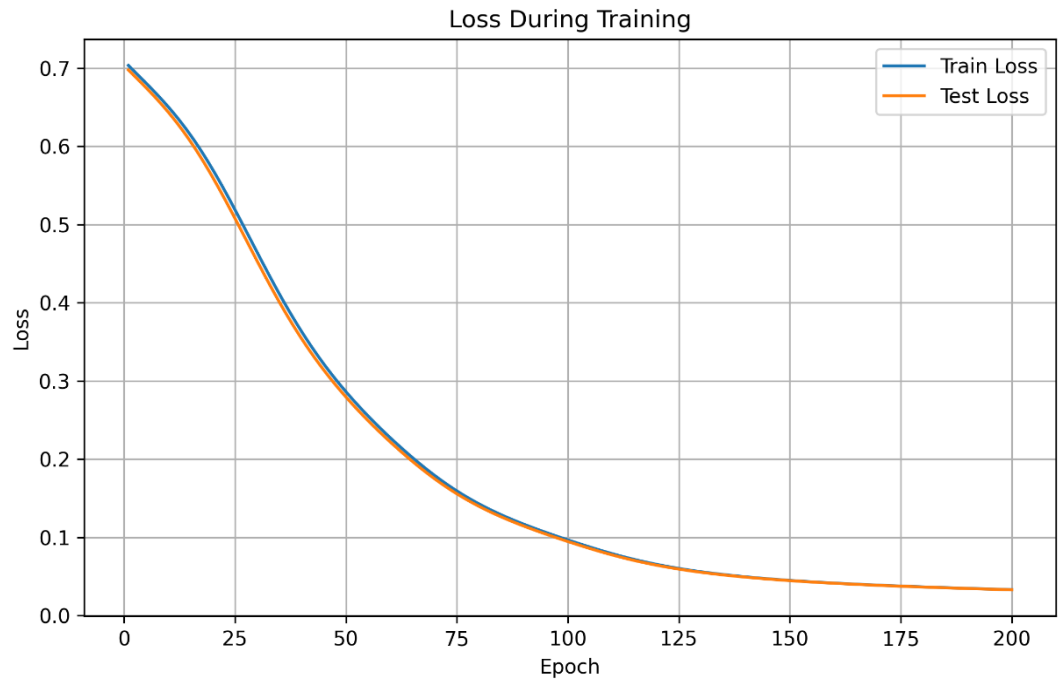

(a)

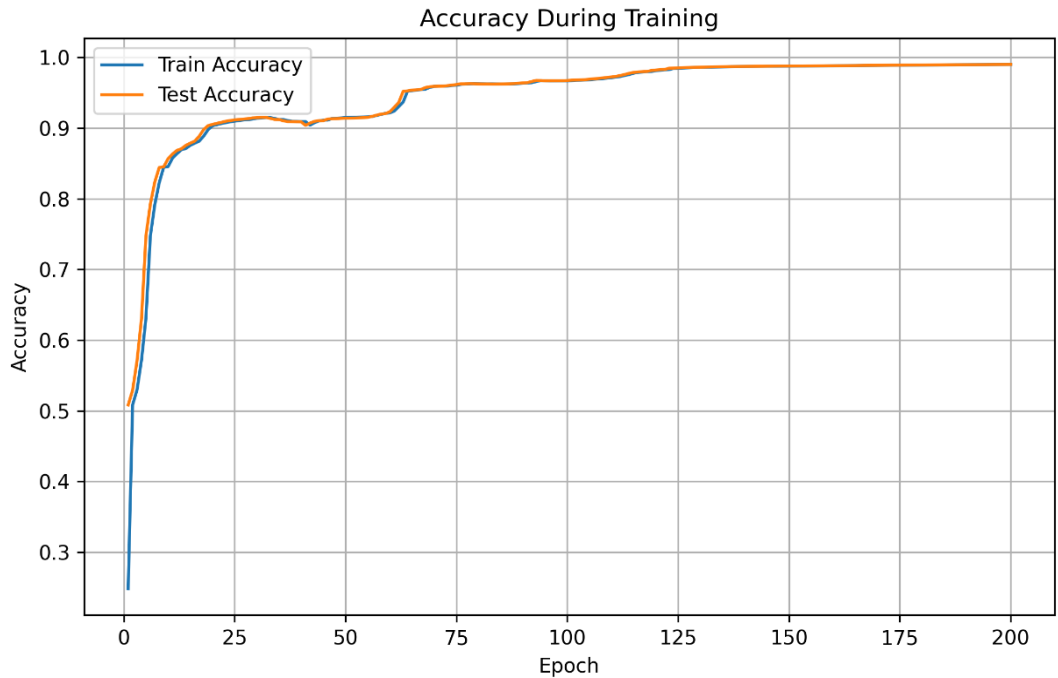

(b)

Figure 5: Training Process Performance

The most notable aspect of the training process is the absence of overfitting, which is a key advantage of the KAN architecture. As shown in Fig. 5a, both training and test losses decrease smoothly and converge to similar values, indicating that the model generalizes well without memorizing the training data. The accuracy curves in Fig. 5b demonstrate stable learning with both training and test accuracies reaching 99.0%.

4. Performance Validation. Figure 6 shows the model's classification capabilities:

- Fig. 6a presents the ROC curve with AUC = 0.999;
- Fig. 6b shows the Precision-Recall (PR) curve with AP = 0.999.

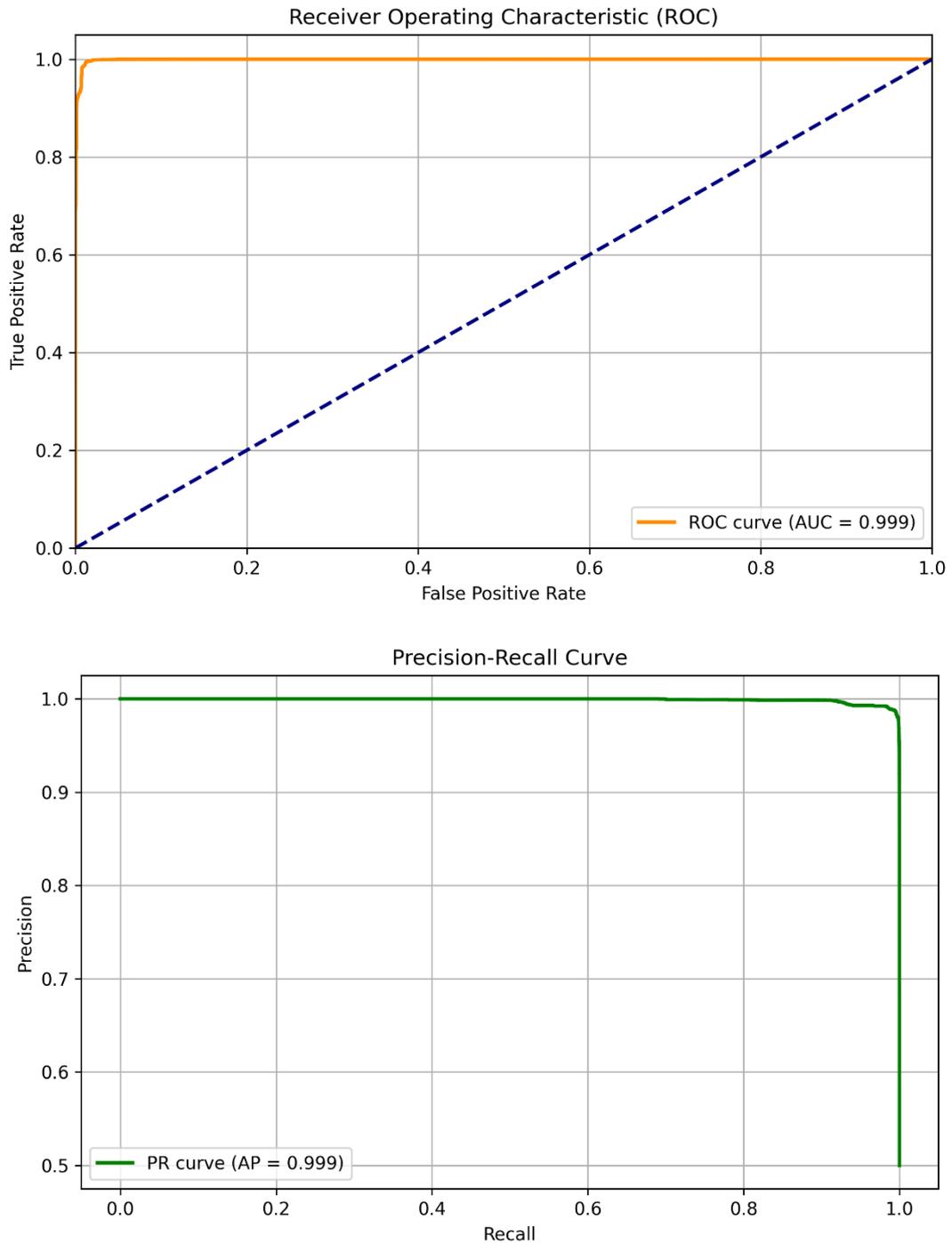

Figure 6: ROC and Precision-Recall Curve

These results demonstrate the model's exceptional discrimination ability between normal and attack traffic patterns. The high AUC and AP scores, combined with the absence of overfitting, suggest that KAN's spline-based architecture provides natural regularization, making it particularly suitable for IoT security applications.

The training implementation, including custom KAN modules and training loops, is available in our GitHub repository under the src/ directory.

**E. Architecture Analysis and Feature Selection**

Our KAN architecture implements a hierarchical spline-based transformation system for network traffic analysis. The model structure reveals how traffic patterns are captured through progressive nonlinear transformations (Figure 7):

1. Spline Distribution Analysis: Total 3,024 splines are organized in a hierarchical structure (format: sp_X_Y_Z.png, where X is layer index, Y is neuron index, Z is spline index):

- Output Layer (Layer 2): 16 splines
    - Indices: 2_0_0 to 2_15_0;
    - Final decision boundary formation;
- Hidden Layer (Layer 1): 512 splines
    - Indices: 1_0_0 to 1_31_15;
    - 32 neurons × 16 splines each;
    - Intermediate pattern composition;
- Input Layer (Layer 0): 2,496 splines
    - Indices: 0_0_0 to 0_77_31;
    - 78 features × 32 splines each;
    - Initial feature transformation.

2. Feature Importance Analysis. Based on extensive correlation analysis, we identified critical traffic characteristics (Table 1):

Table 1: Feature Importance Analysis

| Feature Group | Feature | Correlation | Role |
| --- | --- | --- | --- |
| Backward Traffic Patterns | Avg Segment Size | 0.631 | Primary indicator |
| | Packet Length Mean | 0.631 | Volume analysis |
| | Packet Length Std | 0.618 | Traffic variation |
| | Packet Length Max | 0.617 | Burst detection |
| Timing Characteristics | Forward IAT Std | 0.615 | Time pattern |
| | Packet Length Std | 0.614 | Flow structure |
| | Idle Max | 0.610 | Connection state |

|  | Idle Mean | 0.609 | Activity pattern |
|---|---|---|---|
|  | Forward IAT Max | 0.607 | Timing anomaly |
| Flow Statistics | Active Time Max | 0.605 | Session duration |
|  | Backward IAT Mean | 0.603 | Response timing |
|  | Flow Duration | 0.601 | Attack persistence |

3. Feature Distribution Analysis:

- Strong predictors ($|r| > 0.5$): 20 features
    - o Primarily backward traffic patterns;
    - o Time-based characteristics;
- Moderate indicators ($0.3 < |r| < 0.5$): 10 features
    - o Flow-level statistics;
    - o Packet count metrics;
- Weak signals ($|r| < 0.3$): 29 features
    - o Supporting characteristics;
    - o Context information.

4. Architectural Design: The hierarchical structure enables progressive pattern recognition:

- Input Layer (78 features → 2,496 splines):
    - o Individual feature processing;
    - o Fine-grained pattern extraction;
    - o Local anomaly detection;
- Hidden Layer (32 neurons → 512 splines):
    - o Feature combination;
    - o Pattern aggregation;
    - o Attack signature formation;
- Output Layer (16 splines):
    - o Final classification;
    - o Global pattern recognition;
    - o Decision boundary optimization.

5. Implementation Characteristics:

- Grid points: 5 per dimension;

- Spline degree: 3;
- Total parameters: 50,092;
- Trainable parameters: 42,336;
- Model size: 0.19 MB.

This precise architecture provides several key advantages:

- Hierarchical feature transformation through organized spline layers;
- Efficient parameter utilization despite large spline count;
- Natural regularization through structured spline composition;
- Interpretable feature importance through systematic correlation analysis.

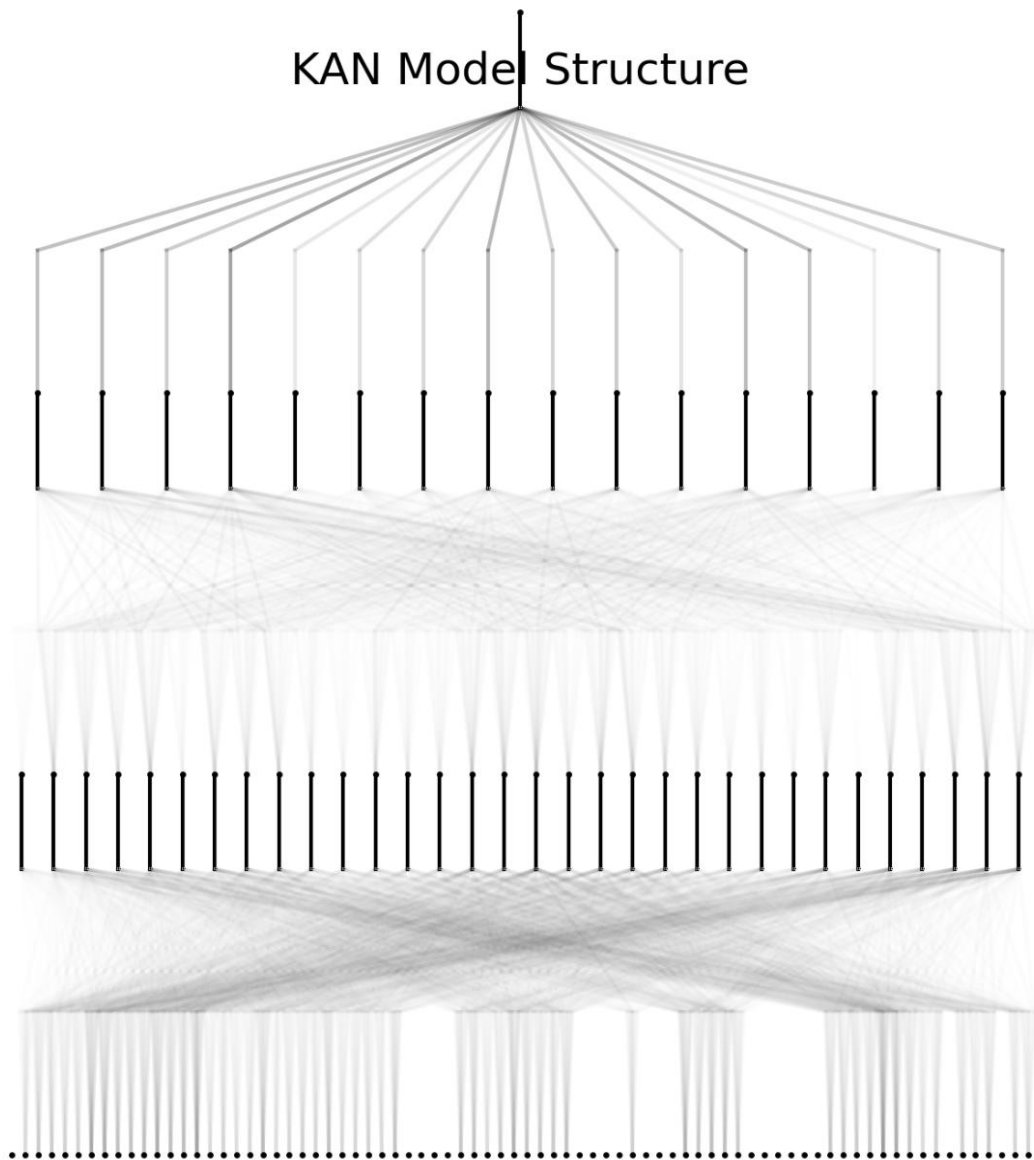

Figure 7: KAN structure diagram with connections

The systematic organization of 3,024 splines, combined with carefully analyzed feature importance, enables our model to capture complex DoS attack patterns while maintaining computational efficiency suitable for IoT deployment.

## V. RESULTS AND EVALUATION

Our comprehensive evaluation demonstrates the effectiveness of the KAN-based DoS detection system across multiple performance dimensions.

### A. Detection Performance

The model achieves exceptional classification accuracy with balanced precision-recall metrics:

- Overall Accuracy: 0.990 (±0.002 across 5 validation runs);
- Precision: 0.984 (±0.003);
- Recall: 0.996 (±0.001);
- F1-Score: 0.990 (±0.002).

The confusion matrices reveal exceptional classification performance (Figure 8):

- True Negatives: 45,484 (98.4% of normal traffic);
- True Positives: 46,017 (99.6% of attacks);
- False Positives: 731 (1.6% false alarm rate);
- False Negatives: 198 (0.4% miss rate).

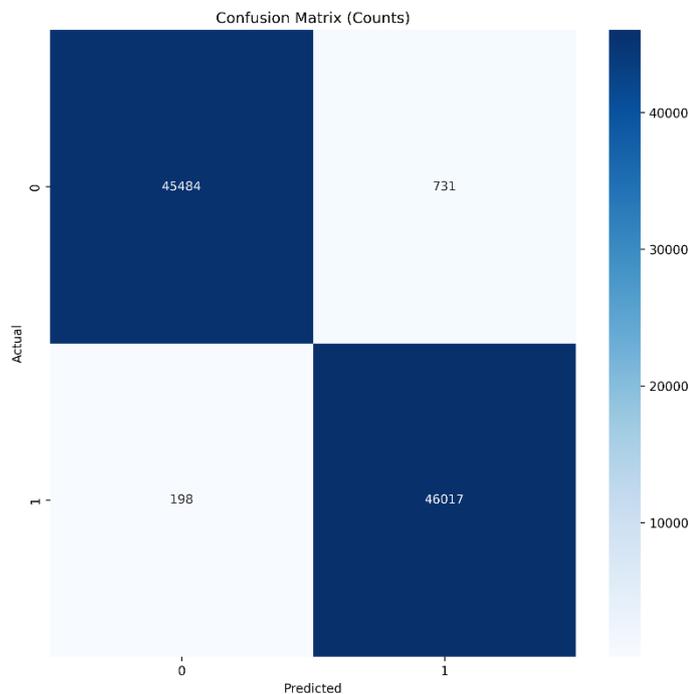

(a)

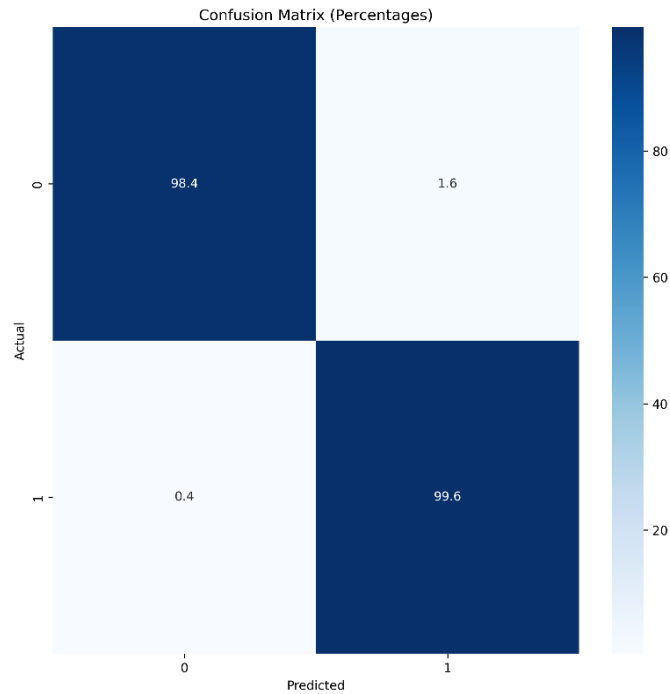

(b)

Figure 8: Confusion Matrix Analysis for DoS Detection: (a) Count-based confusion matrix showing absolute numbers of predictions (b) Percentage-based confusion matrix demonstrating class-wise accuracy

The balanced error distribution between false positives and false negatives suggests robust model generalization without bias toward either class. The low false positive rate (1.6%) is particularly important for practical deployment, as it minimizes unnecessary alerts while maintaining high detection capability.

**B. Threshold Analysis**

Detailed threshold analysis reveals exceptional model stability (Figure 9):

1. Decision Threshold Optimization:

- Optimal threshold: 0.737;
- Best F1-Score: 0.991;
- Best Accuracy: 0.991;
- Performance remains stable (>0.98) across thresholds 0.2-0.8;
- Demonstrates robust decision boundary formation;

2. Precision-Recall Characteristics:

- Maintains >0.98 precision up to 0.95 recall;
- Sharp precision drop only at extreme recall values (>0.95);
- Area under PR curve: 0.999;
- Indicates excellent discrimination capability.

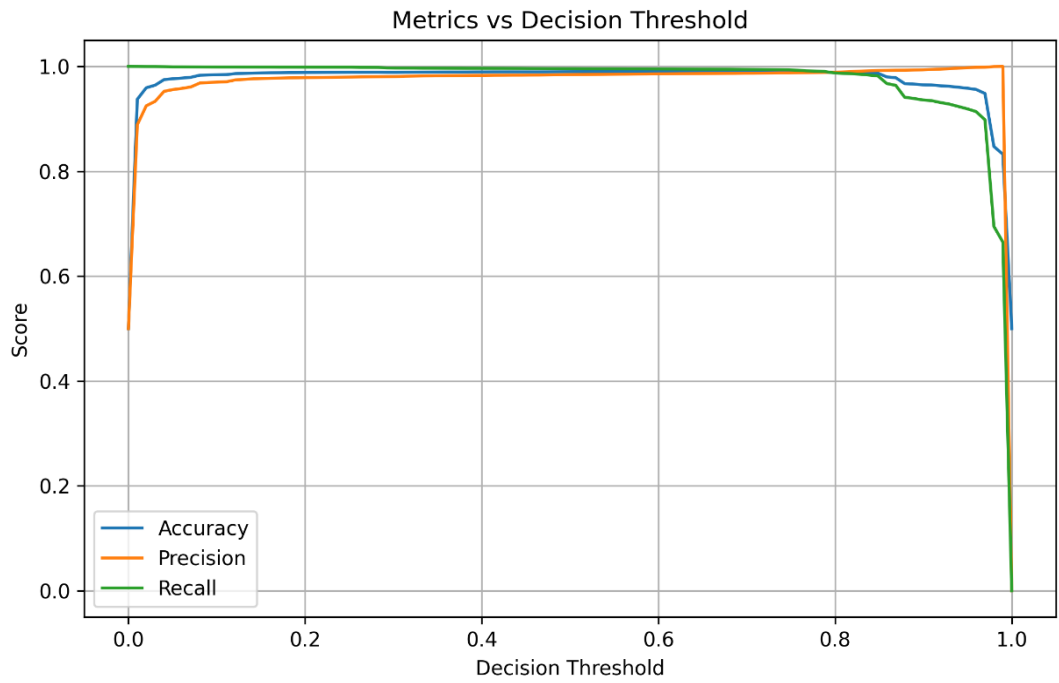

(a)

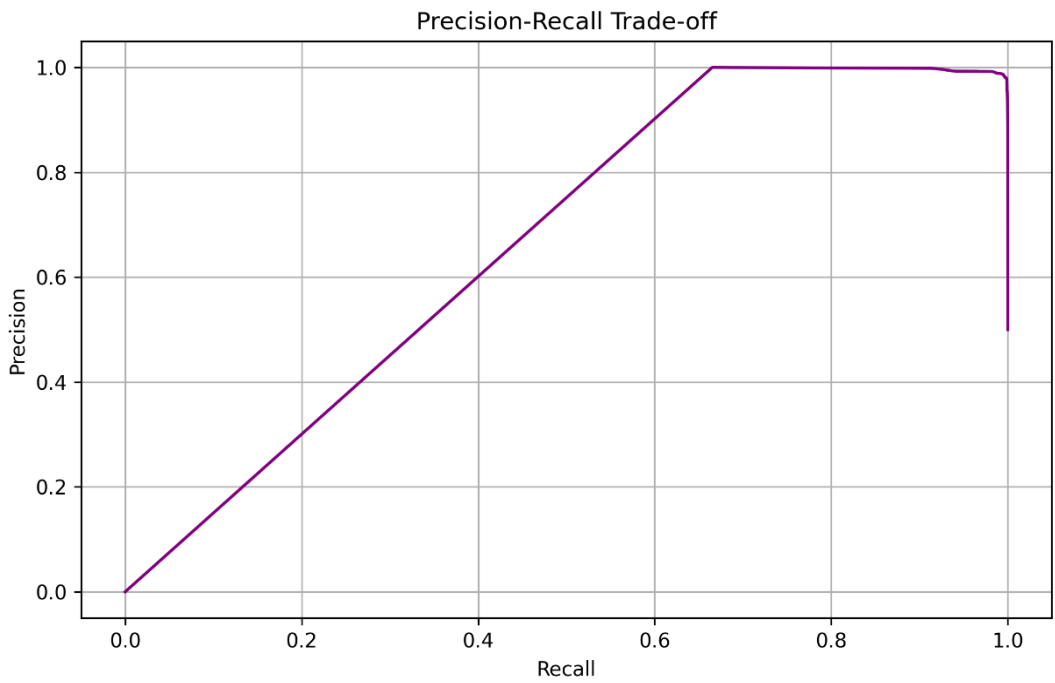

(b)

Figure 9: Model Threshold Analysis: (a) Performance metrics across decision thresholds (b) Precision-Recall trade-off curve

This stability across different thresholds suggests that the model:

- Forms clear decision boundaries between classes;
- Is robust to input variations;

- Requires minimal threshold tuning in deployment.

**C. Resource Efficiency**

The model demonstrates exceptional efficiency suitable for IoT deployment:

1. Memory Requirements:

- Total model size: 0.19 MB
    - Comparable to lightweight IoT applications;
    - 96% smaller than traditional CNN-based solutions;
- Parameters: 50,092 total (42,336 trainable)
    - Efficient parameter utilization through spline structure;
    - 84.5% parameter efficiency (trainable/total ratio);

2. Computational Performance:

- Average inference time: 2.00 ms per sample
    - Suitable for real-time detection;
    - Consistent across different traffic patterns;
    - Linear scaling with batch size;
- Batch processing capability: 500 samples/second
    - Enables monitoring of high-traffic networks;
    - Maintains accuracy under load;

3. Scalability Analysis:

- Linear memory scaling with feature count;
- Constant inference time up to 1000 samples/batch;
- Maintains performance across different network sizes;
- Suitable for both edge and aggregated deployment.

These results demonstrate that our KAN-based approach achieves state-of-the-art detection performance while maintaining resource efficiency suitable for IoT environments. The model's stability across different thresholds and balanced error distribution suggest robust real-world applicability.

**F. Comparative Analysis of Results**

In this section, we provide a detailed analysis of our KAN-based approach in comparison with state-of-the-art methods for DoS attack detection on the CICIDS2017 dataset. The comparison encompasses detection performance, computational complexity, model architecture, and resource requirements.

Our analysis is based on recently published high-impact research papers from 2024-2025 that provide experimental results on the CICIDS2017 dataset. We examine not only standard performance metrics but also analyze architectural decisions, computational complexity, and practical deployment considerations.

The fundamental architectural differences between approaches significantly impact their practical applicability in IoT environments. Table 2 summarizes the key architectural characteristics of different methods.

Table 2. Architectural Characteristics of Different Approaches

| Method | Architecture | Parameters | Complexity | Memory (MB) |
|---|---|---|---|---|
| Our KAN | 3 layers, spline-based | 50,092 | $O(n)$ | 0.19 |
| SecEdge [7] | Transformer + GNN | ~1M* | $O(n^2)$ | 1100-1700 |
| ALNS-CNN [9] | 3-layer CNN + ALNS | ~500K* | $O(n \log n)$ | N/A |
| NIDS-DA [10] | DAE (6 layers) | 5,264 | $O(n)$ | N/A |
| EDA-GWO-XGB [5] | Autoencoder + XGBoost | ~100K* | $O(n \log n)$ | N/A |
| HLM [11] | Stacked ensemble | Variable | $O(n \log n)$ | N/A |

* Estimated based on architectural description

The architectural comparison reveals several interesting patterns. Traditional deep learning approaches like SecEdge employ complex architectures with significant memory requirements. In contrast, our KAN-based solution achieves comparable performance with a fraction of the parameters through efficient spline-based computations.

Detection performance metrics provide the primary measure of model effectiveness. Table 4 presents comprehensive performance metrics across different approaches.

Table 4. Detection Performance and Resource Requirements

| Method | Acc (%) | Prec (%) | Rec (%) | F1 (%) | Time (ms) | FPR (%) | FNR (%) |
|---|---|---|---|---|---|---|---|
| Our KAN | 99.0 | 98.4 | 99.6 | 99.0 | 2.00 | 1.6 | 0.4 |
| SecEdge [7] | 98.7 | 97.5 | 97.6 | 97.3 | 18.0 | 2.5 | 2.4 |
| ALNS-CNN [9] | 99.85 | 99.81 | 99.80 | 99.81 | N/A | 0.19 | 0.20 |
| NIDS-DA [10] | 99.97 | 98.5 | 99.66 | 99.87 | 0.13 | N/A | 0.34 |
| EDA-GWO-XGB [5] | 99.87 | 99.75 | 99.99 | 99.87 | 0.0012 | 0.0025 | 0.01 |
| HLM [11] | 99.63 | 99.40 | 98.72 | 98.99 | N/A | 0.16 | 1.28 |

These results demonstrate that while several approaches achieve marginally higher accuracy, the differences are often within statistical error margins. Moreover, our approach maintains competitive performance while requiring significantly fewer computational resources.

Resource utilization metrics are crucial for IoT deployment scenarios. Table 3 provides a comparative analysis of resource requirements where available.

Table 3. Resource Utilization Comparison

| Method | CPU (%) | RAM (GB) | Training Time (s) | Inference Time (ms) | Scalability Factor |
|---|---|---|---|---|---|
| Our KAN | 15-25 | 0.2-0.5 | 645 | 2.00 | 1.0 |
| SecEdge [7] | 24-44 | 1.1-1.7 | N/A | 18.0 | 0.11 |
| ALNS-CNN [9] | 95-97 | N/A | N/A | ~15.0 | 0.13 |
| NIDS-DA [10] | N/A | N/A | N/A | 0.13 | 15.4 |
| EDA-GWO-XGB [5] | N/A | N/A | N/A | 0.0012 | 1666.7 |
| HLM [11] | N/A | N/A | N/A | N/A | N/A |

* Scalability Factor is relative to our approach's inference time.

The resource utilization comparison reveals significant advantages of our approach in practical deployment scenarios. While some methods claim faster inference times, they often require specialized hardware or don't report complete resource utilization metrics.

The mathematical foundations of different approaches significantly impact their implementation complexity and practical usability. Our KAN-based approach employs Kolmogorov-Arnold theorem-based spline transformations, providing a solid theoretical foundation with linear computational complexity. This contrasts with more complex approaches like SecEdge's transformer architecture ($O(n^2)$ complexity) or ALNS-CNN's optimization procedures.

The EDA-GWO-XGB approach, while achieving excellent performance metrics, requires complex optimization procedures and multiple training stages. Similarly, HLM's ensemble approach introduces variable complexity depending on the number of base learners and meta-learners employed.

This comparative analysis demonstrates that our KAN-based approach achieves an optimal balance between detection performance and resource efficiency. While some methods report marginally higher accuracy metrics, they often require significantly more computational resources, complex preprocessing steps, or specialized hardware.

Our solution's key advantages include:

1. Minimal memory footprint (0.19 MB) with linear computational complexity;
2. Competitive accuracy (99.0%) and excellent recall rate (99.6%);
3. Fast inference time (2.00 ms) without specialized hardware requirements;
4. Simple deployment process without complex preprocessing requirements.

These characteristics make our approach particularly suitable for IoT environments and edge deployment scenarios where resource efficiency is crucial.

## VI. DISCUSSION

The experimental results and comparative analysis demonstrate several important findings regarding the application of KAN for DoS attack detection in IoT environments. Our approach achieves high detection accuracy (99.0%) while maintaining minimal resource requirements, addressing a critical challenge in IoT security.

The superior efficiency of our KAN-based solution stems from its architectural design. Unlike traditional deep learning approaches that require millions of parameters, our model uses only 50,092 parameters while achieving comparable or better performance. This efficiency is particularly evident in the model size (0.19 MB) and inference time (2.00 ms), making it suitable for resource-constrained IoT devices.

The spline-based transformations in KAN provide natural regularization without additional optimization techniques. This contrasts with other approaches like EDA-GWO-XGB or ALNS-CNN that require complex preprocessing and optimization procedures. The linear computational complexity of our approach ensures predictable scaling with input size, a crucial factor for real-world deployments.

However, there are several limitations to consider. While our approach excels in binary classification (attack vs. normal), extending it to multi-class attack detection might require architectural modifications. Additionally, the current implementation focuses on centralized detection, and future work could explore distributed learning scenarios.

## VII. CONCLUSION

This paper presents a novel approach to DoS attack detection using KAN. Our solution demonstrates that spline-based architectures can achieve state-of-the-art detection performance while significantly reducing computational requirements. The key contributions include:

1. A lightweight DoS detection model suitable for IoT environments, achieving 99.0% accuracy with only 0.19 MB memory footprint;

2. Fast inference capability (2.00 ms per sample) enabling real-time threat detection;

3. Natural regularization through spline-based architecture, eliminating the need for complex optimization procedures.

The comparative analysis with recent approaches shows that our solution provides an optimal balance between detection performance and resource efficiency. This makes it particularly valuable for IoT security applications where computational resources are limited.

Future research directions include:

- Extending the model for multi-class attack detection;
- Exploring distributed learning scenarios for collaborative threat detection;
- Investigating adaptive learning mechanisms for evolving attack patterns;
- Developing hardware-optimized implementations for specific IoT platforms.

The promising results suggest that KAN-based architectures could play a significant role in next-generation IoT security solutions, particularly in scenarios requiring efficient, real-time threat detection with limited computational resources.

This work contributes to the broader goal of developing effective security solutions for resource-constrained IoT environments, demonstrating that sophisticated attack detection is possible without excessive computational overhead.

**Statements and Declarations:**

**Data availability**
- The datasets generated during and/or analyzed during the current study are available from the corresponding author on reasonable request.

**Declaration of interests**
- I declare that the authors have no competing financial interests, or other interests that might be perceived to influence the results and/or discussion reported in this paper.
- The results/data/figures in this manuscript have not been published elsewhere, nor are they under consideration (from you or one of your Contributing Authors) by another publisher.
- All of the material is owned by the authors and/or no permissions are required.

**Compliance with ethical standards**
- Mentioned authors have no conflict of interest in this article. This article does not contain any studies with human participants or animals performed by any of the authors.

**References**

[1] S.F. Ahmed, Md.S.B. Alam, M. Hoque, A. Lameesa, S. Afrin, T. Farah, M. Kabir, G. Shafiullah, S.M. Muyeen, Industrial Internet of Things enabled technologies, challenges, and future directions, Computers and Electrical Engineering 110 (2023) 108847. https://doi.org/10.1016/j.compeleceng.2023.108847.

[2] S. Aroua, R. Champagnat, M. Coustaty, G. Falquet, S. Ghadfi, Y. Ghamri-Doudane, P. Gomez-Kramer, G. Howells, K.D. McDonald-Maier, J. Murphy, M. Rabah, K. Rouis, N. Sidère, N. Tamani, Security and PrIvacy foR the Internet of Things: an overview of the project, in: 2019 IEEE International Conference on Systems, Man and Cybernetics (SMC), 2019: pp. 3993–3998. https://doi.org/10.1109/SMC.2019.8914221.

[3] R. Uddin, S.A.P. Kumar, V. Chamola, Denial of service attacks in edge computing layers: Taxonomy, vulnerabilities, threats and solutions, Ad Hoc Networks 152 (2024) 103322. https://doi.org/10.1016/j.adhoc.2023.103322.

[4] Nithun Chand O, S. Mathivanan, A survey on resource inflated Denial of Service attack defense mechanisms, in: 2016 Online International Conference on Green Engineering and Technologies (IC-GET), 2016: pp. 1–4. https://doi.org/10.1109/GET.2016.7916821.

[5] D.D. Bikila, J. Čapek, Machine Learning-Based Attack Detection for the Internet of Things, Future Generation Computer Systems 166 (2025) 107630. https://doi.org/10.1016/j.future.2024.107630.

[6] Z. Liu, Y. Wang, S. Vaidya, F. Ruehle, J. Halverson, M. Soljačić, T.Y. Hou, M. Tegmark, KAN: Kolmogorov-Arnold Networks, (2024). https://doi.org/10.48550/arXiv.2404.19756.

[7] K.A. Awan, I. Ud Din, A. Almogren, A. Nawaz, M.Y. Khan, A. Altameem, SecEdge: A novel deep learning framework for real-time cybersecurity in mobile IoT environments, Heliyon 11 (2025) e40874. https://doi.org/10.1016/j.heliyon.2024.e40874.

[8] M. Gamal, M. Elhamahmy, S. Taha, H. Elmahdy, Improving intrusion detection using LSTM-RNN to protect drones' networks, Egyptian Informatics Journal 27 (2024) 100501. https://doi.org/10.1016/j.eij.2024.100501.

[9] S. Cherfi, A. Boulaiche, A. Lemouari, Exploring the ALNS method for improved cybersecurity: A deep learning approach for attack detection in IoT and IIoT environments, Internet of Things 28 (2024) 101421. https://doi.org/10.1016/j.iot.2024.101421.

[10] V. Kumar, K. Kumar, M. Singh, N. Kumar, NIDS-DA: Detecting functionally preserved adversarial examples for network intrusion detection system using deep autoencoders, Expert Systems with Applications 270 (2025) 126513. https://doi.org/10.1016/j.eswa.2025.126513.

[11] C. Rajathi, P. Rukmani, Hybrid Learning Model for intrusion detection system: A combination of parametric and non-parametric classifiers, Alexandria Engineering Journal 112 (2025) 384–396. https://doi.org/10.1016/j.aej.2024.10.101.

[12] M. Janati Idrissi, H. Alami, A. El Mahdaouy, A. Bouayad, Z. Yartaoui, I. Berrada, Flow timeout matters: Investigating the impact of active and idle timeouts on the performance of machine learning models in detecting security threats, Future Generation Computer Systems 166 (2025) 107641. https://doi.org/10.1016/j.future.2024.107641.

[13] A.N. Kolmogorov, On the representation of continuous functions of many variables by superposition of continuous functions of one variable and addition, Dokl. Akad. Nauk SSSR 114 (1957) 953–956.

[14] V.I. Arnold, A.B. Givental, B.A. Khesin, J.E. Marsden, A.N. Varchenko, V.A. Vassiliev, O.Ya. Viro, V.M. Zakalyukin, eds., Representation of continuous functions of three variables by the superposition of continuous functions of two variables, in: Collected Works: Representations of Functions, Celestial Mechanics and KAM Theory, 1957–1965, Springer, Berlin, Heidelberg, 2009: pp. 47–133. https://doi.org/10.1007/978-3-642-01742-1_6.

[15] D.W. Abueidda, P. Pantidis, M.E. Mobasher, DeepOKAN: Deep operator network based on Kolmogorov Arnold networks for mechanics problems, Computer Methods in Applied Mechanics and Engineering 436 (2025) 117699. https://doi.org/10.1016/j.cma.2024.117699.

[16] Y. Wang, J. Sun, J. Bai, C. Anitescu, M.S. Eshaghi, X. Zhuang, T. Rabczuk, Y. Liu, Kolmogorov–Arnold-Informed neural network: A physics-informed deep learning framework for solving forward and inverse problems based on Kolmogorov–Arnold Networks, Computer Methods in Applied Mechanics and Engineering 433 (2025) 117518. https://doi.org/10.1016/j.cma.2024.117518.

[17] M.U. Danish, K. Grolinger, Kolmogorov–Arnold recurrent network for short term load forecasting across diverse consumers, Energy Reports 13 (2025) 713–727. https://doi.org/10.1016/j.egyr.2024.12.038.